\documentclass[12pt]{article}

\usepackage{latexsym,amsmath,amssymb}
\usepackage{graphicx}
 \usepackage{multicol}
\usepackage[square, sort&compress, numbers, comma]{natbib}

\newcounter{subequation}[equation]
\makeatletter

\def\bcite{\@ifnextchar [{\@tempswatrue\@bcitex}{\@tempswafalse\@bcitex[]}}
\def\@bcitex[#1]#2{\if@filesw\immediate\write\@auxout{\string\citation{#2}}\fi
  \let\@bcitea\@empty
  \@bcite{\@for\@bciteb:=#2\do
    {\@bcitea\def\@bcitea{,\penalty\@m\ }%
     \def\@tempa##1##2\@nil{\edef\@bciteb{\if##1\space##2\else##1##2\fi}}%
     \expandafter\@tempa\@bciteb\@nil
     \@ifundefined{b@\@bciteb}{{\reset@font\bf ?}\@warning
       {Citation `\@bciteb' on page \thepage \space undefined}}%
     \hbox{\csname b@\@bciteb\endcsname}}}{#1}}
\def\@bcite#1#2{{#1\if@tempswa , #2\fi}}

\def\thesubequation{\theequation\@alph\c@subequation}
\def\@subeqnnum{{\rm (\thesubequation)}}
\def\slabel#1{\@bsphack\if@filesw {\let\thepage\relax
   \xdef\@gtempa{\write\@auxout{\string
      \newlabel{#1}{{\thesubequation}{\thepage}}}}}\@gtempa
   \if@nobreak \ifvmode\nobreak\fi\fi\fi\@esphack}
\def\subeqnarray{\stepcounter{equation}
\let\@currentlabel=\theequation\global\c@subequation\@ne
\global\@eqnswtrue
\global\@eqcnt\z@\tabskip\@centering\let\\=\@subeqncr
$$\halign to \displaywidth\bgroup\@eqnsel\hskip\@centering
  $\displaystyle\tabskip\z@{##}$&\global\@eqcnt\@ne
  \hskip 2\arraycolsep \hfil${##}$\hfil
  &\global\@eqcnt\tw@ \hskip 2\arraycolsep
  $\displaystyle\tabskip\z@{##}$\hfil
   \tabskip\@centering&\llap{##}\tabskip\z@\cr}
\def\endsubeqnarray{\@@subeqncr\egroup
                     $$\global\@ignoretrue}
\def\@subeqncr{{\ifnum0=`}\fi\@ifstar{\global\@eqpen\@M
    \@ysubeqncr}{\global\@eqpen\interdisplaylinepenalty \@ysubeqncr}}
\def\@ysubeqncr{\@ifnextchar [{\@xsubeqncr}{\@xsubeqncr[\z@]}}
\def\@xsubeqncr[#1]{\ifnum0=`{\fi}\@@subeqncr
   \noalign{\penalty\@eqpen\vskip\jot\vskip #1\relax}}
\def\@@subeqncr{\let\@tempa\relax
    \ifcase\@eqcnt \def\@tempa{& & &}\or \def\@tempa{& &}
      \else \def\@tempa{&}\fi
     \@tempa \if@eqnsw\@subeqnnum\refstepcounter{subequation}\fi
     \global\@eqnswtrue\global\@eqcnt\z@\cr}
\let\@ssubeqncr=\@subeqncr
\@namedef{subeqnarray*}{\def\@subeqncr{\nonumber\@ssubeqncr}\subeqnarray}
\@namedef{endsubeqnarray*}{\global\advance\c@equation\m@ne%
                           \nonumber\endsubeqnarray}

\renewcommand\maketitle{\par
  \begingroup
    \if@twocolumn
      \ifnum \col@number=\@ne
        \@maketitle
      \else
        \twocolumn[\@maketitle]%
      \fi
    \else
      \newpage
      \global\@topnum\z@   
      \@maketitle
    \fi
    \thispagestyle{plain}\@thanks
  \endgroup
  \setcounter{footnote}{0}%
  \global\let\thanks\relax
  \global\let\maketitle\relax
  \global\let\@maketitle\relax
  \global\let\@thanks\@empty
  \global\let\@author\@empty
  \global\let\@date\@empty
  \global\let\@title\@empty
  \global\let\title\relax
  \global\let\author\relax
  \global\let\date\relax
  \global\let\and\relax
}

\makeatother

\DeclareFontFamily{OT1}{rsfs11}{}
\DeclareFontShape{OT1}{rsfs11}{m}{n}{ <-> rsfs11 }{}
\DeclareMathAlphabet{\mathscript}{OT1}{rsfs11}{m}{n}

\numberwithin{equation}{section}

\newcommand{\pt}{\partial}

\newcommand{\be}{\begin{equation}}
\newcommand{\ee}{\end{equation}}
\newcommand{\nn}{\nonumber}
\newcommand{\bea}{\begin{eqnarray}}
\newcommand{\eea}{\end{eqnarray}}
\newcommand{\bsea}{\begin{subeqnarray}}
\newcommand{\esea}{\end{subeqnarray}}
\newcommand{\hoch}[1]{$\, ^{#1}$}

\def\ie{{\it i.e.\ }}
\def\eg{{\it e.g.\ }}
\def\for{\lower6pt\hbox{$\Big|$}}

\def\ft#1#2{{\textstyle{{\scriptstyle #1}\over {\scriptstyle #2}}}}
\def\fft#1#2{{#1 \over #2}}

\def\G{\Gamma}

\def\O{\Omega}

\def\cD{{\cal D}}

\begin{document}

\begin{titlepage}
\null
\vspace{-1cm}
\begin{flushright}
\small{DAMTP-2007-37 \\
KUL-TF-07/07\\
arXiv:0704.3343 [hep-th]}

\end{flushright}
\begin{center}
\baselineskip=10pt
{\LARGE  Kaluza-Klein Induced Supersymmetry Breaking \\ \vspace{.3cm} for Braneworlds in Type IIB Supergravity}\\
\vspace{10mm}
{\large Jean-Luc Lehners\hoch{\dag}\negthinspace, Paul Smyth\hoch{\ddag} and
K.S. Stelle\hoch{\S} }
\vspace{6mm}

{\small\it \hoch{\dag} DAMTP, CMS, Wilberforce
Road, CB3 0WA, Cambridge, UK \\ \baselineskip=12pt
\smallskip  \hoch{\ddag} Institute for Theoretical Physics, K.U. Leuven,\\   Celestijnenlaan 200D, B-3001 Leuven, Belgium \\ \baselineskip=12pt
\smallskip
\hoch{\S} The Blackett Laboratory, Imperial College London, \\ Prince Consort Road, London SW7 2AZ, UK
}
\vspace{6pt}
\end{center}

\abstract{

We consider $\mathbb{Z}_2$-symmetric braneworlds arising from 5-sphere compactifications with 5-form flux in type IIB
supergravity.  This Kaluza-Klein reduction produces a $D=5$ theory which supports $\frac12$-supersymmetric
$\mathbb{Z}_2$-symmetric domain-wall solutions.  However, upon lifting such solutions back to $D=10$, one finds that
supersymmetry is broken by 5-sphere Kaluza-Klein effects.  This happens owing to the action on the Killing spinor of the
$\mathbb{Z}_2\subset SO(1,9)$ symmetry, which requires an orientation-reversing transformation in the
5-sphere directions  together with the flip of the orbifold coordinate. We study the consequences of this
supersymmetry breaking for the  masses of fermion fluctuation modes about the brane background and find a
natural two-scale hierarchy: some bulk modes  have characteristic masses of order $1\over L_5$ but other
modes more closely associated to the branes have an additional  factor $\exp({-{\rho\over L_5}})$, where
$L_5$ is the $AdS_5$ length parameter and $\rho$ is the orbifold size.}

\vspace{8mm} \vfill \hrule width 2.3cm \vspace{4mm}{\footnotesize
\noindent \hspace{-9mm}
 E-mail: \texttt{j.lehners@damtp.cam.ac.uk, paul.smyth@fys.kuleuven.be, k.stelle@imperial.ac.uk.}
\vskip 5pt
\noindent \hspace{-9mm}\hoch{\dag} Research supported by PPARC.

\noindent \hspace{-9mm}\hoch{\ddag}
Research supported in part by the Belgian Federal Office for Scientific, Technical and \\
\null\hspace{-6mm}Cultural Affairs through the
``Interuniversity Attraction Poles Programme -- Belgian Science Policy"\\
\null\hspace{-6mm}P5/27 and by the EU under contract
MRTN-CT-2004-005104.

\noindent \hspace{-9mm}\hoch{\S}
Research supported in part by the EU under MRTN contract MRTN-CT-2004-005104
and by\\
\null\hspace{-6mm}PPARC under rolling grant PP/D0744X/1.
}

\end{titlepage}
\tableofcontents{}
\setcounter{page}{2}

\section{Introduction}

Perhaps the most dramatic way in which string theory has modified
our picture of spacetime so far is by the inclusion of extra
dimensions and branes. Branes have the remarkable property that
they can localise Yang-Mills gauge theories on them -  this then
leads to the braneworld picture, in which our universe is a brane
embedded in a higher-dimensional bulk spacetime. The
Randall-Sundrum models in particular, in which the bulk is
5-dimensional anti-de Sitter space, have been studied extensively
due to their simplicity and because they provide a possible
solution to the hierarchy problem \cite{Randall:1999ee}, while
also being able to localise gravitons on a brane
\cite{Randall:1999vf}. However, in order to take these models
seriously, one would like to see them emerge as a solution of the
supergravity approximations to string theory. Such a solution was
presented in \cite{Duff:2000az,Cvetic:1999fe} in the context of 5-sphere
compactifications of type IIB supergravity \cite{Bremer:1998zp}
(it is clear from the fact that type IIB supergravity admits an
$AdS_5 \times S^5$ vacuum that this is a natural place to start).
In the present paper we will study this embedding of the
Randall-Sundrum geometry, and the more general family of
$\mathbb{Z}_2$-symmetric braneworlds of which it is a limiting
case, in more detail from a 10-dimensional point of view.

One of the main outstanding questions is how supersymmetry may be broken in
braneworld models, and this is the question that we address here. We find that
supersymmetry is in fact automatically broken at the location of the $\mathbb{Z}_2$
symmetric branes in all the present family of solutions, owing to the geometry of the internal
5-sphere. The reason is the following: due to its chirality (in particular the self-duality of the
5-form), type IIB supergravity actually admits an $SO(1,9)$ rather than an $O(1,9)$ symmetry. This
means that when one is forming a $\mathbb{Z}_2$-symmetric braneworld, one needs to mod out by a
$\mathbb{Z}_2$ that is an element of $SO(1,9)$. Thus, if we flip the orbifold coordinate $y
\rightarrow -y$\,, we must accompany this transformation with a reversal of the orientation of the
internal 5-sphere. However, because of the curvature of the 5-sphere, the Killing spinor equation
is sensitive to the sphere's orientation, and this makes the Killing spinor
discontinuous at the location of the branes. Consequently, supersymmetry is broken on the branes,
while being preserved in the bulk - a phenomenologically attractive setup.

As we will show, this breaking of supersymmetry is not manifest from
the 5-dimensional point of view, but can only be appreciated by
including the internal manifold in the analysis. This is why we call
this type of supersymmetry breaking {\it Kaluza-Klein induced}. It is
also clear from the general argument just presented that this
mechanism will apply to all $\mathbb{Z}_2$-symmetric braneworlds in
type IIB supergravity, as long as the internal manifold is curved.

In order to illustrate the effects of this supersymmetry breaking,
we study a class of bosonic zero modes as well as their fermionic
superpartners. We perform our analysis in linearised perturbation
theory about the braneworld background, taking into account the corresponding
brane actions. The modes that we focus on are those which are
factorisable with regard to their worldvolume and orbifold
dependencies, and which have a profile in the orbifold direction such that, were supersymmetry not
broken, they would appear as massless fields from the 4-dimensional point of view.
Here, however, the fermionic modes acquire a mass, while the
bosons (which are insensitive to the orientation of the 5-sphere)
remain massless. The mass of the fermions depends crucially on
their $y$-dependence. In the most common case, the resulting mass
is naturally of the order of the compactification scale $L_5$, which may be taken to be
near the GUT or Planck scale. However, if the fermionic
modes are such that they have a $y$-dependence that evolves
contrary to the bulk warping, then their mass is suppressed by an additional
bulk warp factor. In this way one obtains two scales of
supersymmetry breaking, and thus both heavy and light fermions, by
the same mechanism.

\section{Dimensional Reduction of Type IIB Supergravity on a 5-Sphere}

Type IIB supergravity can be given a formulation in terms of a
Lagrangian, supplemented by the self-duality condition on the
5-form field. Keeping only the graviton, the gravitino and the
5-form, we have \bea \mathcal{L}_{IIB} &=&
\sqrt{-\hat{g}}\left[\hat{R} - \frac{1}{4\cdot 5!} \hat{F}_{[5]}^2
 - \hat{\bar{\psi}}_M\hat{\G}^{MNP}\hat{\mathcal{D}}_N \hat{\psi}_P\right] \\
\hat{F}_{[5]}&=&*\hat{F}_{[5]}~. \eea Here $\hat{\mathcal{D}}_M$
denotes the supercovariant derivative which also appears in the
supersymmetry transformation of the gravitino: \be
\label{eq:10gravi} \delta \hat\psi_M = \hat{\mathcal{D}}_M
\hat{\epsilon} = \left[\hat\nabla_M +{i\over16\cdot5!}
\hat{F}_{NPQRS}\hat{\G}^{NPQRS}\hat{\G}_M \right]\hat\epsilon\ , \ee
where $\hat{\epsilon}$ is the (chiral) spinorial parameter of the
transformation. We will dimensionally reduce this theory on a
5-sphere $S^5$. In this section we are only interested in the
dimensional reduction of the bulk. We will find a domain wall
solution to this theory, and in the subsequent sections we will be
concerned with the modifications required by the presence of these
lower-dimensional hypersurfaces. The $\G$-matrices are decomposed
according to \bea \hat{\G}^{\underline{m}} &=&
\G^{\underline{m}} \otimes 1 \otimes \sigma_1 \\
\hat{\G}^{\underline{a}} &=& 1 \otimes \tilde\G^{\underline{a}}
\otimes \sigma_2\ , \eea where the $\sigma_i$ are the Pauli matrices.
With this decomposition, the ten-dimensional chirality operator
 is given by $\hat{\G}^{11}=1\otimes 1\otimes \sigma_3$\,. The
10-dimensional fields and the supersymmetry parameter $\hat{\epsilon}$
are then dimensionally reduced according to
\cite{Bremer:1998zp,Liu:2000gk} \bea \label{eq:kka}
ds_{10}^2&=&e^{2\alpha\phi}ds_5^2+e^{2\beta\phi}ds^2(S^5)\\
\hat{F}_{[5]}&=&4me^{8\alpha\phi}\epsilon_{[5]}+4m\epsilon_{[5]}(S^5) \\
\hat{\psi}_m &=& e^{\frac{1}{2}\alpha \phi} (\psi_m + \alpha \G_m \lambda)
\otimes \eta \otimes \left[1\atop 0\right] \\ \hat{\psi}_a &=&
\frac{3 \alpha i}{5} e^{-\frac{11}{10}\alpha \phi} \lambda \otimes \tilde\G_a
\eta \otimes \left[1\atop 0\right] \\ \hat{\epsilon} &=&
e^{\frac{1}{2}\alpha \phi} \epsilon \otimes \eta \otimes \left[1\atop
0\right] \eea where $\phi$ is the breathing mode of the sphere,
\ie $\phi$ determines the volume of the sphere; $\eta$ denotes a
Killing spinor on the 5-sphere. Note that we have chosen
$\hat{\psi}$ and $\hat{\epsilon}$ such that they are of positive
(10-dimensional) chirality. In order to obtain canonically
normalised fields in 5 dimensions, one also has to impose
\be \alpha=\fft14\sqrt{\fft53}\ ,\qquad\beta=-\fft35\alpha\ . \ee
The resulting 5-dimensional bulk theory is the maximal (32-supercharge) $SO(6)$ gauged
supergravity \cite{Kim:1985ez}. The 32-supercharge structure is generated by
four complex, independent 4-component $D=5$ spinors arising from a 5-sphere Killing spinor in the
{\bf 4} of $SU(4) \sim SO(6)$. The $\mathbb{Z}_2$ symmetry we are interested
in acts in the same way on each of these spinors, and so for our purposes it shall
suffice to consider just one of them. Thus, from now on we shall focus on a single
gravitino and adopt a minimal $D=5$ (8 real spinor component) notation (see \cite{Duff:2000jk,Liu:2000gk}). The graviton supermultiplet contains the gravitino $\psi_m$ and a vector. The
breathing mode scalar $\phi$ belongs to a massive vector multiplet
\cite{Kim:1985ez,Liu:2000gk} which also contains a spinor $\lambda$.
Since the two vectors play no role in what follows, we will set
them to zero henceforth. For the reduced set of fields that we
are considering, the 5-dimensional theory is described by the
Lagrangian\footnote{We are not considering higher-order terms in
the fermions here. See also \cite{Gibbons:2000hg} and
\cite{Brito:2001hd} for a discussion of the fermionic equations of
motion.} \bea {\cal L}_5=\sqrt{-g}&&
\hspace{-5mm}\Big[R-\ft12(\partial\phi)^2-V(\phi)-\bar{\psi}_m
\G^{mnp} \mathcal{D}_n \psi_p - \frac{1}{2}\bar{\lambda}\G^m \nabla_m
\lambda -(\frac{1}{4}\frac{\pt^2 W}{\pt \phi^2} - \frac{W}{16})\bar{\lambda}\lambda \nn \\ &&
+\frac{1}{4}\phi_{,m}(\bar{\psi}_m\G^n\G^m\lambda +\bar{\lambda}\G^m\G^n\psi_m) +\frac{1}{8}\frac{\pt W}{\pt
\phi}(\bar{\psi}_m \G^m \lambda - \bar{\lambda}\G^m \psi_m) \Big]\ ,
\label{Lagrangian5d}\eea
where $V(\phi)$ has the double exponential form
\cite{Bremer:1998zp}
\begin{equation}
\label{eq:5pot} V=8m^2e^{8\alpha\phi}-R_5e^{\fft{16}5\alpha\phi}
\end{equation}
and $R_5$ represents the Ricci scalar of $S^5$. The 5-dimensional
supercovariant derivative $\mathcal{D}_n$ is defined in terms of
the superpotential $W$ by \be \mathcal{D}_n = \nabla_n +
\frac{1}{24} W \G_n\ , \ee where
\be
W(\phi) = -8m e^{4 \alpha \phi} + 20 \sqrt{\frac{R_5}{20}}e^{\frac{8}{5}\alpha \phi}\ .
\ee
As usual, the potential $V$ can be written in terms of the
superpotential $W$ according to \be V = \frac{1}{8} \left[W_{,\phi}^2 -
\frac{2}{3}W^2\right]\ . \label{PotFromSuperpot} \ee The fermionic
supersymmetry transformations are \bea
\delta \psi_m &=& \mathcal{D}_m \epsilon = \left(\nabla_m +\frac{1}{24} \G_m W\right)\epsilon \label{gravitino}\\
\delta \lambda &=& \left(\frac{1}{2}\G^m \nabla_m \phi -
\frac{1}{4} W_{,\phi}\right)\epsilon\ . \label{dilatino} \\ \nn
\eea
This 5-dimensional theory admits a two-parameter domain wall
solution given by \cite{Bremer:1998zp} \bea
ds_5^2 &=& e^{2A} dx^\mu dx^\nu \eta_{\mu\nu} + e^{2B} dy^2\ ,\nn\\
e^{-\ft7{\sqrt{15}}\phi} &=& H= -k|y| +c
\ ,\quad B=-4A\ ,\quad k>0\ , \label{DWsolution} \\
e^{4A} &=&  b_1 H^{2/7} +  b_2 H^{5/7}, \nn
\eea
where $ 3k b_1= -28 \, m$ and $3k b_2= + 28 \, \sqrt{\frac{R_5}{20}}$ and $c$ is a
constant. The linear harmonic function $H(y)$ is taken to admit a
second (trough-like) kink at $y=\rho$\,; we thus have a
positive-tension brane at $y=0$ and a negative-tension one at
$y=\rho$ in a Ho\v{r}ava-Witten-like setup with the two branes
located at the endpoints of an $S^1/\mathbb{Z}_2$ orbifold. In the
limit where the scalar $\phi$ sits at the minimum of its potential,
the bulk becomes $AdS_5$ and the double domain wall configuration
then represents the embedding of the Randall-Sundrum model
\cite{Randall:1999ee} in type IIB supergravity \cite{Duff:2000az,Cvetic:1999fe}.

To display the Randall-Sundrum $AdS_5$ patch geometry explicitly \cite{Duff:2000az}, consider the
positive-tension brane at $y=0$ and take $H(0)=c>H_\ast=e^{-{7\over\sqrt{15}}\phi_\ast}+\beta k$,
$\beta>0$, where $\phi_\ast$ is the constant scalar field in the $AdS_5\times S^5$ solution (satisfying
$e^{{24\alpha\over 5}\phi_\ast}={R_5\over20m^2}$). Then take the limit $k\rightarrow 0_+$ and change
coordinates using $\beta-|y|=\beta e^{-{4\over L_5}|z|}$ to obtain the Poincar\'e-coordinate form of
the $AdS_5$ metric
\be
ds^2=e^{-{2|z|\over L_5}}dx^\mu dx^\nu\eta_{\mu\nu}+dz^2\ ,\label{AdS5}
\ee
where
\be
L_5 = m^{-1}\left(20m^2\over R_5\right)^{\frac56} \label{L5}
\ee
is the length parameter of the $AdS_5$ space.

\section{The Supersymmetric Theory in 5 Dimensions}

In order to fully account for the kinks in the double domain wall
solution presented above, we have to extend the $5$-dimensional
theory so as to allow the coupling constants $m$ and $\sqrt{R_5}$ to
change sign when crossing a domain wall. This approach was
developed by Bergshoeff, Kallosh and Van Proeyen (BKvP)
\cite{Bergshoeff:2000zn}, and it allows for a complete
characterisation of $D=5$ supersymmetry, even at the singular brane
hypersurfaces. The easiest way to implement this procedure is to
let \be m \rightarrow m \ \theta(y) \hspace{1cm} \sqrt{R_5}
\rightarrow \sqrt{R_5} \ \theta(y)\ , \ee with \bea \theta(y) &=&
\begin{cases} +1 & \mbox{for } 0 \le y < \pi \\ -1 & \mbox{for } -
\pi \le y < 0
\end{cases}
\eea and we impose the upstairs-picture identification $y \sim
y+2\pi$\,. Note that, consequently, the superpotential should be
redefined as
\be W(y,\phi) = -\theta(y) \left[8m e^{4 \alpha \phi} - 20
\sqrt{\frac{R_5}{20}}e^{\frac{8}{5}\alpha \phi}\right]\ . \ee
Then, the potential of the $5$-dimensional theory can still be expressed by
the relation (\ref{PotFromSuperpot}) and the $5$-dimensional
supersymmetry variations of the gravitino and dilatino remain
unchanged, except for the above new definition of the
superpotential\footnote{The supersymmetric bosonic theory,
including brane actions, has been presented previously in
\cite{Lehners:2005su}.}. The Killing spinor equations, \ie the
vanishing of the above supersymmetry transformations, are then
solved exactly by \be \epsilon=(b_1 H^{2/7}+ b_2 H^{5/7})^{1/8}\epsilon_+\ , \ee
subject to the projection condition $\G_{\underline{y}}\epsilon_+ =
\epsilon_+$\,, where $\epsilon_+$ is a constant spinor. Thus, in the extended
theory with couplings changing sign across domain wall
hypersurfaces, the 5-dimensional braneworld solutions preserve
half of the supersymmetries.

For completeness, and in order to contrast with the
$10$-dimensional calculation that will follow shortly, we
write out the integrability condition for the Killing spinor
equations. Since the Killing spinors have already been found, the
integrability condition is necessarily satisfied; however, it can
be instructive to see the details of how this goes:
\bea
0 &=&
\G^{pmn} [\mathcal{D}_m, \mathcal{D}_n]\epsilon \nn \\ &=& \G^n [{G_n}^p
- \frac{1}{24} {g_n}^p W^2 + \frac{1}{4} {g_n}^p \G^m W_{,m}]\epsilon -
\frac{1}{4} W^{,p}\epsilon\ .
\eea
For $p=y$\,, using $\G_{\underline{y}} \epsilon = \epsilon$\,,
the last two terms cancel, which is in agreement with the fact that there are no singular
contributions to $G_{yy}$\,. For $p=\mu$ the last term vanishes, while the singular terms arising from
$W_{,y}$, where the $y$ derivative acts on the $\theta(y)$ inside $W$,
are cancelled by the singular contributions to the Einstein tensor
\be G_{\mu \nu} = \mathrm{Regular} - g_{\mu \nu} \delta(y)
\frac{1}{\sqrt{g_{yy}}}\left(\frac{3kb_1}{7}H^{-5/7}
+\frac{15kb_2}{14}H^{-2/7}\right)\ .\ee

\section{Oxidising back to 10 Dimensions - Breaking Supersymmetry}\label{d10susybreaking}

The $5$-dimensional domain wall solution can be oxidised back to
10 dimensions, resulting in the metric \cite{Bremer:1998zp} \be
ds_{10}^2 = ( b_1 H^{-3/7} + b_2)^{1/2} dx^\mu dx^\nu\eta_{\mu\nu}
+ ( b_1 H^{13/28} + b_2 H^{25/28})^{-2} dy^2 + H^{3/14} ds^2(S^5)\
. \label{metric10d} \ee Now we can check again the integrability
condition resulting from the Killing spinor equation in 10
dimensions, remembering that we now have $\hat{F}_{[5]} \propto m
\theta(y)$ rather than just $\hat{F}_{[5]} \propto m$\,. We get
\bea 0 &=& {\hat{\G}_P}^{MN} [\hat{\mathcal{D}}_M,
\hat{\mathcal{D}}_N] \epsilon \nn \\ &=& \hat{\G}^N
\left[\hat{G}_{PN} - \frac{1}{96} \hat{F}_{PN}^2\right] \epsilon -
\frac{2i}{4!} \hat{\G}^{QRS} \hat{\nabla}^N \hat{F}_{PQRSN}
\epsilon\ . \label{integrability10d} \eea The bulk terms are
easily seen to satisfy this equation. The singular terms in the
Einstein tensor in 10 dimensions are \cite{Cvetic:2000id}: \bea
\hat{G}_{\mu \nu} &=& \mathrm{Regular} - \hat{g}_{\mu \nu}
\delta(y) \frac{1}{\sqrt{g_{yy}}}
\left(\frac{3kb_1}{7}H^{-15/28} +\frac{15kb_2}{14}H^{-3/28}\right) \label{Einstein10d1} \\ \hat{G}_{yy} &=& \mathrm{Regular} + 0 \\
\hat{G}_{ab} &=& \mathrm{Regular} - \hat{g}_{ab} \delta(y)
\frac{1}{\sqrt{g_{yy}}} \frac{6kb_2}{7}H^{-3/28}
\label{Einstein10d3}\ . \eea For $P=y$\,, the last term in the
integrability condition (\ref{integrability10d}) vanishes, in
agreement with the absence of singular terms in $\hat{G}_{yy}$\,.
For $P=\mu$\,, the last term in the integrability condition adequately
cancels the $b_1$ contribution to the singular terms in
$\hat{G}_{\mu \nu}$; however, there is nothing there to cancel the
singular terms proportional to $b_2$ (and likewise for the $b_2$
terms when $P=a$). We are thus led to conclude that the oxidised domain
wall solution does not preserve any supersymmetry! It is however
supersymmetric away from the branes in the bulk spacetime. Note
that this result also means that the extension of the ordinary
5-dimensional supergravity theory along the lines advocated by
BKvP cannot be obtained by dimensionally reducing type IIB
supergravity.

What, however, is the problem with supersymmetry more concretely? It
is enlightening to study the Killing spinor equations directly;
they are given by the condition \be 0= \delta \hat\psi_M =
\left[\hat\nabla_M +{i\over16\cdot5!}
\hat{F}_{NPQRS}\hat{\G}^{NPQRS}\hat{\G}_M \right]\hat\epsilon\ . \ee
Let us write out this calculation in detail: the $\G$-matrices are
dimensionally reduced to 4+1+5 dimensions according to
\bea
\hat{\G}_{\mu} &=& (b_1 H^{-3/7}+b_2)^{1/4} \gamma_{\mu}\otimes 1 \otimes \sigma_1 \\
\hat{\G}_y &=& (b_1 H^{13/28}+ b_2 H^{25/28})^{-1}
\gamma_y \otimes 1 \otimes \sigma_1
\\\hat{\G}_a &=& H^{3/28} \otimes \tilde{\G}_a \otimes \sigma_2
\eea where there is no $y$-dependence left in $\gamma_{\mu},
\gamma_y, \tilde{\G}_a$ (thus $\gamma_\mu$ are the 4-dimensional
$\G$-matrices with indices raised and lowered with $\eta_{\mu
\nu}$, $\gamma_y$ is the 4-dimensional chirality matrix, and
$\tilde\G_a$ are the $\G$-matrices on the internal 5-sphere).

We are now in a position to analyse the Killing spinor equations,
using $\hat{F}_{[5]} \cdot \hat{\G} = -4i 5!m \theta(y) H^{-15/28}[1
\otimes 1 \otimes (\sigma_1 +i \sigma_2)]$:
\bea
0= \delta\hat{\psi}_{\mu} &=&
\hat{\nabla}_{\mu} + \frac{3kb_1}{112}\theta(y) H^{-15/28}(b_1
H^{-3/7}+b_2)^{1/4}[\gamma_{\mu} \otimes 1 \otimes (1 +\sigma_3)]\hat{\epsilon}
\label{Killing10d1} \\ 0= \delta\hat{\psi}_{y} &=& \hat{\nabla}_{y} +
\frac{3kb_1}{112}\theta(y) H^{-15/28}(b_1 H^{13/28}+b_2
H^{25/28})^{-1}[\gamma_{y} \otimes 1 \otimes (1 +\sigma_3)] \hat{\epsilon} \label{Killing10d2} \\
0= \delta\hat{\psi}_{a} &=& \hat{\nabla}_{a} + \frac{3kb_1}{112}\theta(y)
H^{-15/28} H^{3/28}[1 \otimes \tilde{\G}_a \otimes
i(\sigma_3+1)]\hat{\epsilon}\ , \label{Killing10d3}
\eea
where the spin covariant derivatives are given by
\bea
\hat{\nabla}_{\mu} &=&
\pt_{\mu}
-\frac{3kb_1}{56} \theta(y) H^{-15/28}(b_1 H^{-3/7}+b_2)^{1/4}\gamma_{\mu}\gamma_y \otimes 1 \otimes 1 \\
\hat{\nabla}_{y} &=& \pt_{y} \\ \hat{\nabla}_{a} &=& \nabla_a -\frac{3ik}{56}\theta(y)(b_1
H^{-3/7}+b_2)\gamma_y \otimes \tilde{\G}_a \otimes \sigma_3\ .
\eea

From the expression for the oxidised metric (\ref{metric10d}), we
expect the Killing spinor to be of the form \be \hat{\epsilon} =
(b_1 H^{-3/7} + b_2)^{1/8} \epsilon_+ \otimes \eta \otimes
\left[1\atop 0\right]\ , \ee with $\gamma_y \epsilon_+ =
\epsilon_+$\,. This Ansatz indeed solves the first two Killing
spinor equations (\ref{Killing10d1}) and (\ref{Killing10d2}), but
(\ref{Killing10d3}) reduces to \be \nabla_a \eta = \theta(y)
\frac{i}{2}\sqrt{\frac{R_5}{20}}\tilde{\G}_a \eta\ .
\label{KillingSphere} \ee Using the explicitly known expressions
for Killing spinors on spheres \cite{Lu:1998nu}, we can write down
a solution to the above equation as \be \eta = \eta(y,\theta_a)=
e^{\frac{i}{2}\theta(y) \theta_5\, \frac{3kb_2}{28}
\tilde{\G}_5}\, \Big(\prod_{j=1}^{4} e^{-\ft12 \theta_j\,
(\frac{3kb_2}{28})^2 \tilde{\G}_{j,j+1}} \Big)\, \eta_0\
,\label{sol1} \ee where $\eta_0$ is a constant spinor and
$\theta_a$ with $a,b,...=1,2,3,4,5$ are the angular coordinates on
the 5-sphere. Thus, we see that we are forced to introduce
$y$-dependence into the spherical part of the candidate Killing
spinor in the form of a $\theta$-function. This makes the angular
dependence in the spherical part discontinuous and in this way
supersymmetry is necessarily broken, because a Killing spinor must
be continuous. Note that the change in sign in the angular
dependence corresponds to a reversal of orientation of the
5-sphere. Thus we obtain the geometrical picture that the
orientation of the 5-sphere changes as we cross a brane. This is
consistent with the fact that the type IIB theory admits an
$SO(1,9)$ symmetry rather than an $O(1,9)$ symmetry (this is
because the self-duality of the 5-form must be preserved). Indeed,
the $\mathbb{Z}_2$ symmetry by which we are modding out at the
location of the branes, must be contained within $SO(1,9)$, and
therefore the flip $y \rightarrow -y$ must be accompanied by a
reversal of orientation of the five-sphere.\footnote{Note that for
odd-dimensional spheres such an orientation-reversing map admits
at least two fixed points. From (\ref{sol1}) one sees that our
choice of $\mathbb{Z}_2$ has a fixed point set on the locus
$\theta_5=0$, \ie at the equator of the 5-sphere. Note also that
the antipodal map is orientation-preserving for odd-dimensional
spheres, so this cannot be used as the $S^{5}$ part of the
$\mathbb{Z}_2$ action.}

We should note that there also exists a supersymmetric limit, namely
$b_2 \rightarrow 0$. In this limit the troublesome term
proportional to $\theta(y)$ disappears in the Killing spinor equation (\ref{KillingSphere})
of the sphere, and in fact that condition reduces simply to the condition of
having a covariantly constant spinor. However, this limit is really the
decompactification limit in which the sphere becomes larger and larger, as well as flatter and
flatter, and one ends up with an ordinary 3-brane in 10 dimensions.

Another aspect of the decompactification/supersymmetry-restoring
limit $b_2 \rightarrow 0$ is the structure of the metric warp
factor. In Poincar\'e coordinates (where the transverse term is
simply $dz^2$), the $D=5$ metric in the $b_2 \rightarrow 0$ limit
has a power-law warp factor: \be ds_{5}^{2}=\left(1-{5k\over
7|z|}\right)^{1\over5}dx^{\mu}dx^{\nu}\eta_{\mu\nu}+dz^{2}\
.\label{powerlawwarp} \ee This should be compared with the
structure of the metric in the Randall-Sundrum limit  \cite{Duff:2000az} $k\rightarrow0$\,,
where the Poincar\'e-coordinate metric (\ref{AdS5}) is
composed of patches of anti-de Sitter space with an exponential
warp factor: \be ds_{5}^{2}=e^{-{2|z|}\over
L_{5}}dx^{\mu}dx^{\nu}\eta_{\mu\nu}+dz^{2}\
.\label{exponentialwarp} \ee The exponential warp factor underlies
many of the proposed physical applications of the Randall-Sundrum
schemes, be it the effective concentration of gravity near the
$D=4$ positive tension brane in RSII, or possible applications
to the hierarchy problem arising from exponential differences in coupling
constants on opposing RSI braneworlds. These features disappear with
the power-law warp factor (\ref{powerlawwarp}) which arises as
supersymmetry is restored in the $b_2 \rightarrow 0$ limit.

Let us take stock at this point of what we have learned: we have a
double domain wall solution in 5 dimensions, which upon oxidation
to 10 dimensions on a 5-sphere leads to another double domain wall
solution. This solution has the particular property that it is
supersymmetric everywhere in the bulk spacetime, but breaks
supersymmetry completely at the locations of the domain walls. This
immediately raises two questions:

1. Since supersymmetry is broken on the domain wall, what is the
mass scale of this breaking, as seen from the viewpoint of a
4-dimensional observer on the domain wall? We will treat this
question in the section \ref{Scale}.

2. Is this solution stable? Indeed one might speculate that since
this non-supersymmetric solution is surrounded by a supersymmetric
spacetime, it might be kept stable by the surrounding bulk (in the
fully supersymmetric case solutions of this type are known to be
stable despite the presence of a negative-tension brane
\cite{Lehners:2005su}). A detailed calculation of the stability
properties of this solution would be very interesting. We leave it
for future work.

\section{Fermionic Modes and the Scale of Supersymmetry Breaking} \label{Scale}

If supersymmetry were not broken, we would expect the theory on
the 4-dimensional branes to be an ungauged supergravity theory
with half the number of supercharges as compared with the bulk theory \cite{Duff:2000jk}. Then,
there would be fermionic modes which, from the 4-dimensional point
of view, would be massless. In this section, we will present modes
of this type, which we obtain as superpartners of linearised
massless bosonic perturbations (see the Appendix for a detailed
derivation, following the linearised supersymmetry procedure of
\cite{Cvetic:2002su}). However, since supersymmetry is actually broken, we
know that the fermionic excitations will pick up mass terms (while
the bosons remain massless at this level). The easiest way to derive these mass
terms is by dimensionally reducing the 10-dimensional
Rarita-Schwinger action for these modes in order to find their
4-dimensional effective actions. The mass terms then arise when a
$y$-derivative hits the discontinuity in the spherical spinor part
$\eta(y,\theta_a)$ at the location of the branes\footnote{The setup
described here thus provides a concrete example of the
general framework for brane supersymmetry breaking of Bagger and Belyaev
\cite{Bagger:2002rw,Bagger:2004rr}.}.

\subsection{The Gravitino}

As shown in the Appendix, one of the would-be massless
perturbation modes of the braneworld geometry that we are
considering is a worldvolume
gravitino given by \bea \psi_\mu &=& (b_1 H^{2/7} + b_2 H^{5/7})^{1/8} \tilde\psi_\mu (x) \\ \psi_y &=& 0 \\
\lambda &=& 0\ . \eea This mode can be lifted to 10 dimensions, where it
reads \bea \hat{\psi}_{\mu} &=& (b_1 H^{2/7} + b_2 H^{5/7})^{1/8}
\tilde\psi_{\mu} \otimes \eta(y,\theta_a) \otimes \left[1\atop 0\right] \\
\hat{\psi}_y &=& 0 \\ \hat\psi_a &=& 0\ . \eea The discontinuous
spherical part of the gravitino should really be seen as an
approximation; one would expect the gravitino to be continuous
but interpolating between two different bulk profiles on either side of a brane.
For our purposes, though, this approximation is
accurate enough. The action for the gravitino can then be
dimensionally reduced as follows\footnote{We define
$\bar{\psi}=\psi^{\dagger}A$\,. The 10-dimensional intertwiner
$A_{1,9}$ is dimensionally reduced according to $A_{1,9} = A_{1,4}
\otimes A_{0,5} \otimes \sigma_2$\,. Our conventions are as in Sohnius
\cite{Sohnius:1985qm}.}
\bea
&&\int_{10d} \sqrt{-\hat{g}}i \, \hat{\bar{\psi}}_M \hat{\G}^{MNP}\hat{\mathcal{D}}_N \hat{\psi}_P\  = \
\int_{4d} \sqrt{-g} \int_0^{\rho} H^{-5/14}dy \int d\O_4
\int_0^{\pi} \sin^4(\theta_5)d\theta_5 \times \nn \\
&&
\ \left(i (b_1 H^{-3/7}+b_2)^{-1/2} \bar{\tilde\psi}_{\mu} \gamma^{\mu \nu
\tau} \nabla_{\nu} \tilde\psi_{\tau} \otimes \bar{\eta}
  \eta \otimes \left[1 \ 0\right] \sigma_2 \sigma_1 \left[1\atop 0\right]\right. \nn \\
&&
\ \left.-i(b_1 H^{2/7}+b_2 H^{5/7})^{3/4} H^{5/14}i
\frac{3kb_2}{28} \theta_5
\left(\delta(y)-\delta(y-\rho)\right)\bar{\tilde\psi}_{\mu} \gamma^{\mu \nu} \gamma_y
\tilde\psi_{\nu} \otimes \bar{\eta} \tilde{\gamma}_5 \eta \otimes \left[1 \
0\right] \sigma_2 \sigma_1 \left[1\atop 0\right]\right)\ . \nn
\eea
Now, if we assume that $\tilde{\gamma}_5 \eta = \pm \eta$ (thus the $SO(6)$
symmetry of the five-sphere also gets broken at the location of the
branes), and further use the fact that $\gamma_y \psi_{\mu} =
\psi_{\mu}$ as well as the integrals
\bea \int_0^{\pi} \sin^4(\theta_5) d\theta_5 &=& \frac{3\pi}{8} \\
\int_0^{\pi} \sin^4(\theta_5) \theta_5 d\theta_5 &=& \frac{3\pi^2}{16}\ , \eea then we get the 4-dimensional effective action for
$\tilde\psi_\mu:$ \bea S_{\tilde\psi} = \int_{4d} \sqrt{-g} [i \,
\bar{\tilde\psi}_{\mu} \gamma^{\mu\nu\tau}\nabla_{\nu}\tilde\psi_{\tau} \pm
m_{(3/2)} \ \bar{\tilde\psi}_{\mu} \gamma^{\mu \nu} \tilde\psi_{\nu}]\ , \eea where \be m_{(3/2)} = \frac{3\pi k b_2}{56} \frac{[(b_1
H^{2/7}+b_2 H^{5/7})^{3/4}]_0^{\rho}}{\int_0^{\rho}dy (b_1 H^{2/7}+b_2
H^{5/7})^{-1/2}}\ . \ee Thus, as expected, we find an ungauged
supergravity in 4 dimensions, broken by the above mass term.

The expression for the mass term is a bit unwieldy, which is why it is instructive
to write out the Randall-Sundrum limit of the above formulae.  The 10-dimensional
metric is then given by \be ds^2_{10,RS} = e^{-\frac{2|y|}{L_5}}\eta_{\mu
\nu}dx^{\mu}dx^{\nu}+dy^2+L_5^2d\O_5^2\ , \ee where $L_5$ is the length parameter
of both $AdS_5$ and $S^5$ as given in (\ref{L5}). The gravitino reduces to
\bea
\psi_\mu &=& e^{-\frac{|y|}{2L_5}} \tilde\psi_\mu (x) \\ \psi_y &=& 0 \\
\lambda &=& 0\ ,
\eea
which in 10 dimensions is expressed as \bea \hat{\psi}_{\mu}
&=& e^{-\frac{|y|}{2L_5}} \tilde\psi_{\mu}(x) \otimes \eta(y,\theta_a) \otimes
\left[1\atop 0\right] \\
\hat{\psi}_y &=& 0 \\ \hat\psi_a &=& 0\ . \eea In the Randall-Sundrum limit the
discontinuous spherical spinor $\eta$ is given by \be \eta(y,\theta_a)=
e^{\frac{i}{2}\theta(y) \theta_5\,
\tilde{\gamma}_5}\, \Big(\prod_{j=1}^{4} e^{-\ft12 \theta_j\,
\tilde{\gamma}_{j,j+1}} \Big)\, \eta_0\ . \ee The calculation of the effective
action proceeds along the same lines as above, and this time we find \bea
S_{\tilde\psi, RS} = \int_{4d} \sqrt{-g} [i
\, \bar{\tilde\psi}_{\mu} \gamma^{\mu\nu\tau}\nabla_{\nu}
\tilde\psi_{\tau} \pm m_{(3/2,RS)} \ \bar{\tilde\psi}_{\mu}
\gamma^{\mu \nu} \tilde\psi_{\nu}]\ , \eea where \be m_{(3/2,RS)} =
\frac{\pi}{L_5}
\frac{(1-e^{-\frac{3\rho}{L_5}})}{(1-e^{-\frac{2\rho}{L_5}})}\ .
\ee
We can see that $m_{(3/2,RS)}$ varies between
$\frac{3\pi}{2L_5}$ (as $\frac{\rho}{L_5} \rightarrow 0$) and
$\frac{\pi}{L_5}$ (as $\frac{\rho}{L_5} \rightarrow \infty$), and is therefore
always close to the $L_5^{-1}$ scale of
$S^5$-compactification, which one may take to be close to the GUT or Planck
scales. Thus, from the 4-dimensional point of view, the gravitino is heavy.

\subsection{Other Modes}

Let us now turn our attention to the other fermionic modes
discussed in the Appendix. These modes have a different
$y$-profile, but nevertheless, in a supersymmetry-preserving
context, they would appear to be massless from a 4-dimensional
perspective. First of all, we have the fermionic partner of the
Goldstone boson associated with the $y$-translation symmetry that is
broken by the brane. This mode is given by \bea \psi_\mu &=&
-\frac{k}{8H} {\gamma_\mu}^\rho s_{,\rho} (2 b_1 H^{2/7} + 5 b_2 H^{5/7})
(b_1 H^{2/7} + b_2 H^{5/7})^{-3/8} \epsilon_+
\\ \psi_y &=& -\frac{k}{4H} (2 b_1 H^{2/7} + 5 b_2 H^{5/7}) (b_1
H^{2/7} + b_2 H^{5/7})^{-13/8} \gamma^\rho s_{,\rho} \epsilon_+ \nn \\ &&+
\frac{k^2 \theta(y)}{112 H^2}
(b_1 H^{2/7} + b_2 H^{5/7})^{-11/8} [24 b_1^2 H^{4/7} + 60 b_1 b_2 H
+ 45 b_2^2 H^{10/7}] s(x) \epsilon_+ \\
\lambda &=& \frac{\sqrt{15}k}{2H} (b_1 H^{2/7} + b_2 H^{5/7})^{3/8}
\gamma^\rho s_{,\rho} \epsilon_+\ , \eea which in the Randall-Sundrum limit
simplifies to \bea \psi_\mu &=& \frac{1}{4} e^{\frac{3|y|}{2L_5}}
{\gamma_{\mu}}^\rho s_{,\rho} \epsilon_+
\\ \psi_y &=& \frac{1}{2} e^{\frac{5|y|}{2L_5}} \gamma^{\rho} s_{,\rho} \epsilon_+
-\frac{\theta(y)}{2L_5}e^{\frac{3|y|}{2L_5}} s(x) \epsilon_+
 \\ \lambda &=& 0\ . \eea
If we now let \be {\gamma_{\mu}}^\rho s_{,\rho} \epsilon_+ \equiv \gamma_\mu \chi \ee and take this as the
definition of the mode $\chi(x)$, then we can derive the
effective 4-dimensional action for $\chi:$ \bea S_{\chi, RS} = \int_{4d} \sqrt{-g} [i \,
\bar{\chi} \gamma^{\mu}\nabla_{\mu} \chi \pm m_{(\chi,RS)} \ \bar{\chi}
\chi]\ , \eea where the mass term is given by \be m_{(\chi,RS)} =
\frac{6\pi}{5L_5}
\frac{(e^{\frac{\rho}{L_5}}-1)}{(e^{\frac{2\rho}{L_5}}-1)}\ . \ee We can see
that this time $m_{(\chi,RS)}$ varies between
$\frac{3\pi}{5L_5}$ (as $\frac{\rho}{L_5} \rightarrow 0$) and
$e^{-\frac{\rho}{L_5}} \frac{6\pi}{5L_5}$ (as $\frac{\rho}{L_5} \rightarrow
\infty$). Thus we get an exponential mass suppression when
$\frac{\rho}{L_5}$ is large. It seems reasonable on phenomenological
grounds to assume that $\rho$ might be an order of magnitude larger than
$L_5$ \cite{Randall:1999ee}, and therefore $\chi$ can be a {\it light}
fermion from the 4-dimensional effective theory point of view. This is
because $\chi$ has a profile along the orbifold direction $y$ which
evolves in the opposite way as compared to the bulk warp factor, and
therefore $\chi$ is localised mainly near the negative-tension
brane at $y=\rho$\,.

The last mode that we will consider is the fermionic partner to the
third bosonic mode presented in the Appendix. This bosonic mode
has the particular property that in the Randall-Sundrum limit it
reduces to a pure scalar field perturbation, the metric remaining
unchanged. In general, its fermionic partner is given by \bea
\psi_\mu &=& -\frac{k}{8H^{10/7}}
{\gamma_\mu}^\rho s_{,\rho} (b_1 H^{2/7} + b_2 H^{5/7})^{17/8} \epsilon_+ \\
\psi_y &=& -\frac{k}{4 H^{10/7}}  (b_1 H^{2/7} + b_2
H^{5/7})^{7/8} \gamma^\rho s_{,\rho} \epsilon_+ \nn \\ && + \frac{3 k^2 b_1
\theta(y)}{28 H^{15/7}}
(b_1 H^{2/7} + b_2 H^{5/7})^{9/8}  s(x) \epsilon_+ \\
\lambda &=& \frac{\sqrt{15}k}{8H} (b_1 H^{2/7} + b_2 H^{5/7})^{-1/8} [2
b_1^2 H^{1/7} + b_1 b_2 H^{4/7} - b_2^2 H] \gamma^\rho s_{,\rho} \epsilon_+\ , \eea which in the Randall-Sundrum limit reduces to a pure dilatino
perturbation: \bea \psi_\mu &=& 0
\\ \psi_y &=& 0 \\ \lambda &=& \frac{1}{2} e^{-\frac{7|y|}{2L_5}} \gamma^{\rho} s_{,\rho}
\epsilon_+\ . \eea It is again helpful to define $\tilde\lambda$ by \be
\frac{1}{2} \gamma^{\rho} s_{,\rho} \epsilon_+ \equiv \tilde\lambda\ , \ee in terms of
which the effective action is given by \bea S_{\tilde\lambda, RS} =
\int_{4d} \sqrt{-g} [i \, \bar{\tilde\lambda}\gamma^{\mu}\nabla_{\mu}
\tilde\lambda \pm m_{(\tilde\lambda,RS)} \ \bar{\tilde\lambda} \tilde\lambda ]\ , \eea
with the mass term \be m_{(\tilde\lambda,RS)} = \frac{5\pi}{L_5}
\frac{(1-e^{-\frac{11\rho}{L_5}})}{(1-e^{-\frac{10\rho}{L_5}})}\ . \ee
$m_{(\tilde\lambda,RS)}$ thus varies between $\frac{11\pi}{2L_5}$ (as
$\frac{\rho}{L_5} \rightarrow 0$) and $\frac{5\pi}{L_5}$ (as
$\frac{\rho}{L_5} \rightarrow \infty$), and so $\tilde\lambda$ is another
example of a heavy fermion.

\section{Discussion}

We have seen that $\mathbb{Z}_2$-symmetric braneworlds in type IIB
supergravity necessarily break supersymmetry owing to the chiral nature of the theory and the
curvature of the internal manifold. Supersymmetry is broken only at the location of the
branes, and this can also be traced back to the presence of source
terms that are proportional to the square root of the curvature
$\sqrt{R_5} \sim b_2$ of the internal 5-sphere.

We have shown how fermionic modes, which would have been massless
in a supersymmetric context, thus acquire masses. Moreover,
depending on their profiles along the orbifold direction $y$, the
effective 4-dimensional fermionic modes can appear either heavy or
light. The heavy modes are those whose profiles along the orbifold
direction evolve similarly to the metric warp factor, \ie they are
mainly associated to the bulk geometry,  and they tend to have a
mass comparable to the $L_{5}^{-1}$ compactification scale. The
light fermions, by contrast, are those modes which have profiles
that evolve in the opposite way as compared to the metric warp factor
and are more specifically associated to a brane. In the
Randall-Sundrum limit, for example, the light fermions are those
that have a $y$-dependence proportional to \be
e^{\frac{c|y|}{L_5}} \qquad \mathrm{with} \qquad c>1\ . \ee This
ensures that these modes are mostly localised near the
negative-tension brane at $y=\rho$\,. For large values of
$\frac{\rho}{L_5}$\,, their masses are suppressed by a factor of
$e^{-\frac{\rho}{L_5}}$\,, which is certainly attractive for
phenomenological reasons. We recall the discussion of section
\ref{d10susybreaking} on the exponential warp factor in the
general solutions of Bremer {\it et al.}\,\cite{Bremer:1998zp}, and
particularly in the Randall-Sundrum limit \cite{Duff:2000jk}, as
compared to the power-law warp factor occurring in the
supersymmetry-restoring $b_2 \rightarrow 0$ limit. This
exponential warp factor is also seen to be at the root of the orbifold
exponential hierarchy of masses for fermionic fluctuations that we
have found.

For simplicity, we have focused in this paper on the minimal
$D=5$ supersymmetric structure with 8 supercharges. The full
$S^5$-reduced theory, of course, has an extended 32-supercharge
supersymmetry organized into a {\bf 4} of $SU(4)\sim SO(6)$. In
the bulk spacetime of the brane solutions we consider, each of
these 4 $D=5$ spinor supercharges splits up into two $D=4$
spinors of opposite $\Gamma_{\underline{y}}$ chirality, one of
which becomes spontaneously broken, with a corresponding
massive gravitino in the usual fashion. The remaining
$\Gamma_{\underline{y}}$ chirality gives the erstwhile unbroken
supersymmetry, which however is broken by the $\mathbb{Z}_2$
structure of the  brane system as we have shown. Choosing a
specific $\mathbb{Z}_2$ action in the 5-sphere directions
necessitates picking an equator of the 5-sphere, which becomes
the fixed-point surface for the chosen $\mathbb{Z}_2$. This
breaks the surviving automorphism symmetry down from $SU(4)\sim
SO(6)$ to $USp(4)\sim SO(5)$. However, the {\bf 4}
representation remains irreducible with respect to $USp(4)$, so
all 4 of the $D=5$ theory's supersymmetries get broken by the
$\mathbb{Z}_2$ action in the same way. Accordingly, the full
story is just a four-fold replication of the minimal $D=5$
story that we have presented.

Let us conclude with a few remarks about the nature of the
supersymmetry-breaking sources. The Randall-Sundrum scenario has
been associated to a combination of D3 branes and  7-branes
\cite{Chan:2000ms}. There is a possible association of the $b_2$ term
in our construction to 7-branes, as noted already in \cite{Cvetic:2000id}.
Note that the singular terms in $G_{ab}$ are a factor of
$\frac{4}{5}$ smaller than the ones in $G_{\mu \nu}$, suggesting
that the upper 4-dimensional parts of the worldvolumes of the
7-branes might be averaged over the 5 spherical dimensions. This
association is supported furthermore by the fact that the
$y$-dependence of the singular terms in
(\ref{Einstein10d1})-(\ref{Einstein10d3}) would be consistent with
the presence of two transverse directions instead of just one, \eg
the $y$-direction and one of the spherical directions. For the
$b_2$ part of the solution, we explicitly have \be G_{\mu \nu}
\sim G_{ab} \sim b_2 \delta(y) (b_1 H^{5/14} + b_2 H^{11/14}) =
b_2 \frac{\delta(y)}{\sqrt{g_{\mathrm{transverse}}}}\ , \ee where
(with no summation implied on $a$) \be g_{\mathrm{transverse}} =
g_{yy}g_{aa}\ . \ee Note also that the $\mathbb{Z}_2$ symmetry
chosen in our construction has a fixed-point set on the locus
$y=0\ , \theta_{5}=0$, \ie an 8-dimensional surface which could be
associated to a 7-sphere worldvolume.

Going against the 7-brane interpretation of the $b_{2}$ part of the solution,
however, is the fact that the IIB axionic scalar that would support a standard
7-brane is zero in the background considered here. Of course it could be that the
precise smearing of 7-branes needed has to be such that the axion charge averages
to zero.\footnote{Another puzzle with such an interpretation arises in the
analogous case of 11-dimensional supergravity compactified on a 7-sphere. In that
case, the analogous source would have to be made out of 8-branes, but no 8-brane
solutions are known in D=11 supergravity, although they do exist in massive type
IIA supergravity \cite{Romans:1985tz}.}

A final question is that of stability. Even if the background
solution we consider can be associated to a smeared set of
7-branes taken together with the D3 brane, the breaking of supersymmetry
that we have found raises the question of whether this construction has tachyonic
instabilities. But since the bulk spacetime remains perfectly
supersymmetric away from the branes, one is led to speculate that
the bulk supersymmetry might be enough to stabilise the boundary
branes where supersymmetry is broken, perhaps in a manner similar
to the ``fake supergravity'' framework of Ref.
\cite{Freedman:2003ax}.

\section*{Acknowledgments}

The authors would like to thank Jussi Kalkkinen for collaboration
during the early stages of this work. They would also like to
acknowledge useful and stimulating discussions with Ben Allanach,
Kevin Costello, Mirjam Cveti\v{c}, Ruth Durrer, Joel Fine, Gary
Gibbons, Ulf Gran, Neil Lambert, Jim Liu and Antoine Van Proeyen.
K.S.S. would like to thank the TH Unit at CERN and the Galileo
Galilei Institute for Theoretical Physics for hospitality and would like to
thank the INFN for partial support during the completion of this work.

\section*{Appendix: Linearised Domain Wall Perturbation Modes}

\addcontentsline{toc}{section}{Appendix: Linearised Domain Wall
Perturbation Modes}

\renewcommand{\theequation}{\Alph{section}.\arabic{equation}}
\setcounter{section}{1} \setcounter{equation}{0}

In this Appendix, we explicitly derive the form of linearised
bosonic perturbations about domain wall geometries, which, from
the 4-dimensional point of view, are massless. We also show how
one can then determine the fermionic superpartners of these modes.
It should be noted that the method employed here is not the same
as determining the moduli of a domain wall solution and then
promoting those moduli to spacetime-dependent fields (see for
example \cite{Lehners:2006ir} for an exposition of the latter
method). Here, we allow the various modes to have different
$y$-dependent profiles along the orbifold direction, chosen such
that the modes appear massless from the brane worldvolume
perspective (when supersymmetry is not broken). The existence of
this type of zero mode is really a particular feature of
braneworld Kaluza-Klein reductions. Consider theories of the form
\be S=\int_{5d}
\sqrt{-g}\left[R-\frac{1}{2}(\partial\phi)^2-V(\phi)\right] -
\int_{4d, y=0}\sqrt{-g}W(\phi)\ , \ee where, in this Appendix, we
are considering a single positive tension domain wall residing at
$y=0$. In static gauge, the equations of motion are \bea G_{mn}
&=&
\frac{1}{2}\phi_{,m}\phi_{,n}-\frac{1}{4}g_{mn}\phi_{,p}\phi^{,p}-
\frac{1}{2} g_{mn} V(\phi) - \frac{1}{2}
\frac{\delta(y)}{\sqrt{g_{yy}}}\delta^{\mu}_m \delta^{\nu}_n
g_{\mu \nu}
W(\phi)\\
\Box{\phi} &=& \frac{\pt V}{\pt \phi} +\frac{\delta(y)}{\sqrt{g_{yy}}}
\frac{\pt W}{\pt \phi}\ .
\eea
We write the fields as \bea g_{mn} &=& g_{mn}^{(0)}+h_{mn} \\
\phi &=& \phi^{(0)} + \phi^{(1)}\ ,
\eea
where $^{(0)}$ quantities
correspond
to the unperturbed domain wall solutions. We then have
\bea g^{mn} &=& g^{mn(0)} - h^{mn} \\
\G^{(1)p}_{mn} &=& \frac{1}{2}
g^{(0)pl}(h_{lm;n}+h_{ln;m}-h_{mn;l})\ . \eea When we perturb the
geometry, we choose coordinates such that the domain wall always
remains at $y=0$ \cite{Charmousis:1999rg}. The linearised
equations of motion then are \bea &&
\frac{1}{2}({h^p}_{m;np}+{h^p}_{n;mp}-{h_{mn;p}}^p-h_{;mn})
-\frac{1}{2}g^{(0)}_{mn}({h^{pl}}_{;lp}-{h_{;p}}^p) \nn \\ &&
-\frac{1}{2}h_{mn}R^{(0)}
+\frac{1}{2}g_{mn}^{(0)}h^{pl}R_{pl}^{(0)} =
\frac{1}{2}\phi^{(0)}_{,m}\phi^{(1)}_{,n}+\frac{1}{2}\phi^{(0)}_{,n}\phi^{(1)}_{,m}
 -\frac{1}{4}h_{mn}\phi^{(0),p}\phi^{(0)}_{,p} \nn \\ &&
+\frac{1}{4}g^{(0)}_{mn}h^{pl}\phi^{(0)}_{,p}\phi^{(0)}_{,l}
-\frac{1}{2}g^{(0)}_{mn}\phi^{(0),p}\phi^{(1)}_{,p}
-\frac{1}{2}h_{mn}V -\frac{1}{2}\frac{\pt V}{\pt \phi}\phi^{(1)}\nn \\
&& - \frac{1}{2}
\frac{\delta(y)}{\sqrt{g^{(0)}_{yy}}}\delta^{\mu}_m \delta^{\nu}_n
\left(h_{\mu \nu} W+g^{(0)}_{\mu \nu}\frac{\pt W}{\pt \phi}\phi^{(1)} -
\frac{1}{2}{h^y}_y g^{(0)}_{\mu \nu} W\right)\eea and \bea && \Box^{(0)}
\phi^{(1)} -h^{mn}\phi^{(0)}_{;mn} -{h^{mn}}_{;n}\phi^{(0)}_{;m}
+\frac{1}{2}h^{,m}\phi^{(0)}_{,m} \nn
\\ &&= \frac{\pt^2 V}{\pt \phi^2}\phi^{(1)} +
\frac{\delta(y)}{\sqrt{g^{(0)}_{yy}}} \left(\frac{\pt^2W}{\pt
\phi^2} \phi^{(1)} - \frac{1}{2}{h^y}_y \frac{\pt W}{\pt
\phi}\right)\ . \eea For a background metric of the form \be ds^2
= e^{2A(y)} \eta_{\mu \nu} dx^{\mu} dx^{\nu} + e^{2B(y)}dy^2\ ,
\ee the non-zero connections are \be \G^{(0)y}_{\mu \nu} =
-A_{,y}\eta_{\mu \nu} e^{2A-2B} \qquad \G^{(0)\rho}_{\mu y} =
\delta^{\rho}_{\mu} A_{,y} \qquad \G^{(0)y}_{yy} = B_{,y}\ , \ee
and thus we have \bea R^{(0)}_{\mu \nu} &=& \eta_{\mu \nu}
e^{2A-2B} (-A_{,yy} -4A_{,y}^2 + A_{,y} B_{,y})
\\ R^{(0)}_{yy} &=& -4A_{,yy} -4A_{,y}^2 + 4A_{,y} B_{,y}
\\ G^{(0)}_{\mu \nu} &=& \eta_{\mu \nu} e^{2A-2B} (3A_{,yy} +6A_{,y}^2 -3 A_{,y} B_{,y})
\\ G^{(0)}_{yy} &=& 6A_{,y}^2 \\
\Box^{(0)} \phi^{(0)} &=& e^{-2B}(\phi^{(0)}_{,yy} + 4
\phi^{(0)}_{,y}A_{,y} - \phi^{(0)}_{,y} B_{,y})\ . \eea Taking into
account that we are imposing a $\mathbb{Z}_2$ symmetry at the
location of the domain wall, we can see that the junction
conditions (\ie the matching conditions for the singular pieces in the Einstein
equations) at the location of the domain wall become
\bea
12A_{,y} &=& - e^B W \quad |_{y=0} \\
2 \phi^{(0)}_{,y} &=& e^B \frac{\pt W}{\pt \phi} \quad |_{y=0}\ .
\eea
Note that the $yy$ Einstein equation is given by \be 6A_{,y}^2 =
\frac{1}{4}\phi^{(0)2}_{,y} - \frac{1}{2}e^{2B}V\ . \ee
This doesn't involve second derivatives in $y$\,, which is consistent with the
fact that there are no singular source terms in that direction. If
we evaluate this equation at the location of the domain wall, we
can substitute in the junction conditions derived above, to find
\be
V = \frac{1}{8} \left(\left(\frac{\pt W}{\pt \phi}\right)^2 - \frac{2}{3}
W^2\right)\ \for_{y=0}\ .
\ee
This is of course the relation between the
superpotential $W$ and the potential $V$ in supersymmetric
theories. In a supersymmetric context, the junction conditions
above are actually the Bogomol'nyi equations, and they are then
valid throughout the bulk. Furthermore, this shows that the domain
wall couples to the bulk via the superpotential.

Looking at the terms containing two $y$ derivatives in the
linearised equations of motion, one can write down the linearised
junction conditions in this background (making use of the 0th
order junction conditions):
\bea
(h_{\mu \nu} - \eta_{\mu
\nu}\eta^{\rho \sigma} h_{\rho \sigma})_{,y} &=& -\frac{1}{6}e^B W (h_{\mu
\nu} - \eta_{\mu \nu}\eta^{\rho \sigma} h_{\rho \sigma})
+\frac{1}{4}e^{2A-B}\eta_{\mu \nu} W h_{yy} \nn \\ &&
+\frac{1}{2} e^{2A+B}\eta_{\mu \nu} \frac{\pt W}{\pt \phi}\phi^{(1)} \ \for_{y=0}
\label{linearjct1}\\
\phi^{(1)}_{,y} &=& \frac{1}{2} e^{B} \frac{\pt^2 W}{\pt
\phi^2}\phi^{(1)} +\frac{1}{4} e^{-B} \frac{\pt W}{\pt \phi} h_{yy}
\ \for_{y=0}\ . \label{linearjct2}
\eea
There is also a junction
condition associated with the $\mu y$ linearised Einstein
equation, and it reads
\be
h_{\mu y} = 0 \ \for_{y=0}\ .
\ee
This condition was already implied by the imposition of the
$\mathbb{Z}_2$ symmetry at $y=0$ under which $h_{\mu y}$ is odd.

\subsection*{Examples of Bosonic Modes}

\addcontentsline{toc}{subsection}{Examples of Bosonic Modes}

We will now give explicit expressions for these modes in the
Bremer {\it et al.} case \cite{Bremer:1998zp} as well as the
Randall-Sundrum limit \cite{Duff:2000az,Cvetic:1999fe}.

In the Bremer {\it et al.} case \cite{Bremer:1998zp}, the
background solution is given by \be e^{-\ft7{\sqrt{15}}\phi} = H=
-k|y| +c \ ,\qquad e^{4A} =  b_1 H^{2/7} + b_2
H^{5/7}\ , \qquad B=-4A\  .\ee We then have the following
expressions for the superpotential and the potential: \bea W &=&
\frac{3k}{7}(2b_1 H^{-5/7}+5b_2 H^{-2/7})\theta(y)
\\ V &=& \frac{9k^2}{196}(2b_1^2 H^{-10/7}-5b_2^2
H^{-4/7})\ . \eea The $\mu \neq \nu$ linearised junction
conditions are solved for $h_{\mu \nu} \sim (b_1 H^{2/7} + b_2
H^{5/7})^{1/2}$\,, unless $h_{\mu \nu} \propto \eta_{\mu \nu}$\,. In
the first case, this ansatz also solves the other junction
conditions and the linearised equations of motion (it should be
noted that, although the linearised junction conditions would also allow
for $h_{yy}$ and $\phi^{(1)}$ contributions, say
proportional to a mode $c(x)$, the $\mu \neq \nu$ linearised Einstein
equations would then demand $c_{, \mu \nu} = 0$ and thus we set
$h_{yy} = 0 = \phi^{(1)}$). This mode represents a 4-dimensional
worldvolume graviton excitation: \bea h_{\mu \nu} &=& (b_1 H^{2/7}
+ b_2 H^{5/7})^{1/2} \tilde{h}_{\mu \nu} (x)
\\ h_{yy} &=& 0 \\
\phi^{(1)} &=& 0\ , \eea where $\tilde{h}_{\mu \nu} (x)$ obeys the
4-dimensional linearised Einstein equations.

In the second case, our ansatz for the metric perturbations is
\bea
h_{\mu \nu} &=& a \eta_{\mu \nu} s(x^{\rho}) f(y) \\ h_{yy} &=&
s(x^{\rho}) j(y) \ .
\eea
Then, looking at the linearised $\mu \nu$
equations for $\mu \neq \nu$ we find
\bea
a &=& -\frac{1}{2} \\
j(y) &=& (b_1 H^{2/7} + b_2 H^{5/7})^{-5/2} f(y)\ .
\eea
Note that these conditions also automatically ensure that there are no
$s_{;\mu \nu}$ terms present in $G^{(1)}_{\mu \nu}$ for any
$\mu,\nu$. Next we look at the $\mu y$ equations, from which we
can infer that
\be \phi^{(1)} =
\frac{21\theta(y)}{2\sqrt{15}k}s(x^{\rho})(b_1 H^{2/7} + b_2
H^{5/7})^{-1/2}H f_{,y}\ .
\ee
At this point all $\mu m$ equations
are identically satisfied for all $f(y)$. The linearised $yy$ and
$\phi$ equations demand \be \Box^{(4d)}s(x^\rho)=0 \ee and the
additional constraint
\bea
0 &=& f_{,yy} [98 b_1^2 H^{4/7}+98
b_2^2 H^{10/7}+196 b_1 b_2 H] \nn \\ &&
+k f_{,y} [154 b_1^2 H^{-3/7} +91 b_2^2 H^{3/7}+245 b_1 b_2 ] \nn \\
&& + k^2 f [-10 b_1^2 H^{-10/7}-10 b_2^2
H^{-3/7}-20 b_1 b_2 H^{-1}] \label{Constraintyy}\ .
\eea
This has the following two solutions:
\bea f(y) &=& 2 b_1
H^{-5/7} + 5 b_2 H^{-2/7}
\qquad \propto (b_1 H^{2/7}+b_2 H^{5/7})_{,y} \\
f(y) &=& (b_1 H^{-2/7} +b_2 H^{1/7})^{5/2}\ .
\eea
Thus, explicitly, we have the ``Goldstone'' mode \bea h_{\mu \nu} &=& -\frac{1}{2}
\eta_{\mu \nu} s(x^{\rho})(2 b_1 H^{-5/7}+5 b_2 H^{-2/7})k
\\ h_{yy} &=& s(x^{\rho})(b_1 H^{2/7}+b_2 H^{5/7})^{-5/2}(2 b_1 H^{-5/7}+5 b_2 H^{-2/7})k \\
\phi^{(1)} &=& s(x^{\rho}) \sqrt{15}(b_1 H^{2/7}+ b_2
H^{5/7})^{1/2}H^{-1}k \eea and a third mode \bea
h_{\mu \nu} &=& -\frac{1}{2} \eta_{\mu \nu} s(x^{\rho})(b_1
H^{-2/7}+ b_2 H^{1/7})^{5/2}k
\\ h_{yy} &=& s(x^{\rho})H^{-10/7} k \\
\phi^{(1)} &=& \frac{\sqrt{15}}{4} s(x^{\rho})(b_1 H^{2/7}+
b_2 H^{5/7})(2 b_1 H^{-8/7}- b_2
H^{-5/7})k\ . \eea It is then straightforward to verify that
the latter two modes also satisfy the linearised junction
conditions. The Goldstone mode takes its name from the fact that, in
the bulk,
its general form can be obtained by a $y$-dependent diffeomorphism with
parameter
\bea
\xi_\mu &=& 0 \\ \xi_y &=& 7 (b_1 H^{2/7} + b_2 H^{5/7})^{-3/2}s\ ,
\eea where
\bea
h_{mn} &=& \xi_{m;n} + \xi_{n;m} \\ \phi^{(1)} &=& \xi^m \phi_{,m}\ . \eea If we then promote $s$ to a function
$s(x)$\,, this is not a diffeomorphism anymore, and we obtain the
above non-trivial mode. In this sense, this mode is a Goldstone mode
corresponding to the translational symmetry that is broken by the
domain wall (see \cite{Adawi:1998ta} for a general treatment of
these types of modes).

For the Randall-Sundrum model \cite{Randall:1999ee,Duff:2000az}, we have the
following expressions for the superpotential and the potential:
\bea && W=\frac{12}{L_5}\theta(y) \qquad \frac{\pt W}{\pt \phi}=0 \qquad
\frac{\pt^2 W}{\pt \phi^2}=-\frac{8}{L_5}\theta(y) \\ &&
V=-\frac{12}{L_5^2} \qquad \frac{\pt V}{\pt \phi}=0 \qquad \frac{\pt^2
V}{\pt \phi^2}=\frac{32}{L_5^2}\ , \eea where $L_5$ is the $AdS_5$ radius of
curvature\footnote{Incidentally, the second derivative of the
potential indicates that the breathing mode $\phi$ has mass
squared equal to $32/L_5^2$\,, in agreement with \cite{Kim:1985ez}.}. The
background metric is given by \be ds_5^2 =
e^{-\frac{2|y|}{L_5}}\eta_{\mu \nu}dx^{\mu}dx^{\nu} + dy^2\ . \ee The
perturbation modes can simply be determined by taking the
appropriate limit of the Bremer {\it et al.} modes
\cite{Duff:2000az}. This gives the graviton excitation
 \bea h_{\mu \nu} &=&
e^{-\frac{2|y|}{L_5}} \tilde{h}_{\mu \nu} (x)
\\ h_{yy} &=& 0 \\
\phi^{(1)} &=& 0\ , \eea where $\tilde{h}_{\mu \nu} (x)$ obeys the
4-dimensional linearised Einstein equations. The Goldstone mode is
now given by \bea h_{\mu \nu} &=& s(x) \eta_{\mu \nu}
\\ h_{yy} &=& -2e^{\frac{2|y|}{L_5}} s(x) \\
\phi^{(1)} &=& 0\ , \eea with \be \Box^{(4d)}s(x)=0\ . \ee This is the
``radion'' mode of Ref. \cite{Charmousis:1999rg} (see also
\cite{Lehners:2002tw} for a heterotic M-theory equivalent), and it
can be obtained by starting with a diffeomorphism with parameter
$\xi_y = -\frac{L_5}{2}e^{\frac{2|y|}{L_5}}s$\,. The third mode reduces
to a pure scalar field perturbation: \bea h_{\mu \nu} &=& 0
\\ h_{yy} &=& 0 \\
\phi^{(1)} &=&  e^{-\frac{4|y|}{L_5}} s(x)\ , \eea again with
$\Box^{(4d)}s(x)=0$.

\subsection*{Fermionic Partners}

\addcontentsline{toc}{subsection}{Fermionic Partners}

The fermionic superpartners of the bosonic modes that we have just
derived can be obtained by using the linearised form of the
supersymmetry transformations (\ref{gravitino},\ref{dilatino}) \cite{Cvetic:2002su}. In
this way it is guaranteed that the resulting fermions are also
solutions of the linearised equations of motion. In general, the
fermions are given by \bea \psi_m &=& (\cD)^{(1)}_m \epsilon \\ \lambda &=&
\left[-\frac{1}{4}{h^y}_y \gamma^y \phi_{,y} +\frac{1}{2} \gamma^\mu
\phi^{(1)}_{,\mu} + \frac{1}{2} \gamma^y \phi^{(1)}_{,y} - \frac{1}{4}
\frac{\pt^2 W}{\pt \phi^2} \phi^{(1)}\right]\epsilon\ , \eea with \bea
(\cD)^{(1)}_\mu &=& \frac{1}{4}h_{\mu \nu, \rho} \gamma^{\nu \rho}
 + \frac{1}{4}(h_{\mu \nu ,y} -A_{,y} h_{\mu \nu}
 -g_{\mu \nu} A_{,y} {h^y}_y)\gamma^{\nu y} \nn \\ && +\frac{W}{48} {h_\mu}^\nu \gamma_\nu +\frac{1}{24} \frac{\pt
W}{\pt \phi} \phi^{(1)}\gamma_\mu \\
(\cD)^{(1)}_y &=& -\frac{1}{4} h_{yy,\mu} \gamma^{\mu y} +\frac{W}{48}
{h_y}^y \gamma_y +\frac{1}{24} \frac{\pt W}{\pt \phi} \phi^{(1)}\gamma_y\ . \eea
In the case of a graviton perturbation \be h_{\mu \nu} = e^{2A}
\tilde{h}_{\mu \nu}(x) \qquad h_{yy} = 0 = \phi^{(1)}\ , \ee the
fermionic superpartner is the gravitino given by \bea \psi_\mu &=&
\frac{1}{4} h_{\mu \nu,\rho} \gamma^{\nu \rho} e^{\frac{A}{2}} \epsilon_+
= \frac{1}{4} \tilde{h}_{\mu \nu,\rho} \gamma^{\nu \rho} e^{\frac{A}{2}} \epsilon_+
\equiv e^{\frac{A}{2}}\tilde\psi_\mu(x) \\ \psi_y &=& 0 \\
\lambda &=& 0\ . \eea Consequently the chirality of the gravitino is
given by \be \gamma_y \tilde\psi_\mu = + \tilde\psi_\mu\ . \ee
Our remaining bosonic modes are of the form \bea h_{\mu \nu} &=&
-\frac{1}{2} \eta_{\mu \nu} s(x) f(y) \\ h_{yy} &=& s(x) e^{2B-2A}
f(y)\ . \eea In this case the linearised junction conditions
(\ref{linearjct1}) and (\ref{linearjct2}) simplify the resulting
expressions for the fermionic partners, and we end up with
\bea
\psi_\mu &=& -\frac{1}{8} e^{-\frac{3}{2}A} f {\gamma_{\mu}}^\rho s_{,\rho} \epsilon_+
\\ \psi_y &=& -\frac{1}{4} e^{-\frac{5}{2}A+B} f {\gamma^{\mu}}  s_{,\mu} \epsilon_+
+\frac{W}{48} e^{-\frac{3}{2}A+B} f s(x) \epsilon_+ +\frac{1}{24} \frac{\pt W}{\pt \phi} \phi^{(1)}e^{\frac{A}{2}+B}
\epsilon_+ \\
\lambda &=& \frac{1}{2} \gamma^\mu \phi^{(1)}_{,\mu} e^{-\frac{A}{2}} \epsilon_+\ .
\eea
Thus we can see that we have the following chiralities \bea \gamma_y \psi_\mu &=& + \psi_\mu \\
\gamma_y \lambda &=& - \lambda\ , \eea whereas $\psi_y$ contains terms of both
chiralities.

As a consistency check, it is straightforward to verify that all
the above fermionic modes satisfy their equations of motion: \bea
&& \gamma^{mnp} \cD_n \psi_p -\frac{1}{8} \frac{\pt W}{\pt \phi} \gamma^m \lambda - \frac{1}{4} (g^{mn} -
\gamma^{mn})\phi_{,n} \lambda = 0 \label{EomPsi} \\
&& \gamma^m \nabla_m \lambda + (\frac{1}{2}\frac{\pt^2 W}{\pt
\phi^2}-\frac{W}{8})\lambda - \frac{1}{2} \gamma^m \gamma^n \phi_{,n} \psi_m +
\frac{1}{4} \frac{\pt W}{\pt \phi} \gamma^m \psi_m = 0\ . \label{EomLambda} \eea


\end{document}